\documentstyle[twocolumn,aps,floats,psfig]{revtex}
\begin{document}
\author{Mauricio Cataldo%$^a
\thanks{On leave of absence from the Universidad del B\'\i o-B\'\i o,
Concepci\'on,
Chile}%^
 {\thanks{E-mail address: mcataldo@alihuen.ciencias.ubiobio.cl}}%$
and Alberto Garc\'\i a%$^a
\thanks{
On sabattical leave from Departamento de F\'\i sica CINVESTAV--IPN, Apartado
Postal 14--740, C.P. 07000, M\'exico, D.F. M\'exico}
 \thanks{E-mail address: agarciad@lauca.usach.cl}%$
}
\address{%$^a$
Departamento de F\'\i sica, Facultad de Ciencia,
Universidad de Santiago de Chile, Avda. Ecuador 3493, Casilla 307,  Santiago,
Chile.
%\\ $^b$Departamento de F\'\i sica, Facultad de Ciencia,
%Universidad de Santiago de Chile, Avda. Ecuador 3493, Casilla 307, Santiago,
%Chile.
\\ \smallskip\ }
\title{Three Dimensional Black Hole Coupled to the Born-Infeld
Electrodynamics.}
\maketitle
\begin{abstract}
{\bf Abstract:}{A nonlinear charged version of the (2+1)-anti de Sitter black
hole solution is derived. The source to the Einstein equations is a Born-Infeld
electromagnetic field, which in the weak field limit becomes the
(2+1)-Maxwell field. The obtained Einstein-Born-Infeld solution  for certain
range of the parameters (mass, charge, cosmological and the Born-Infeld
constants) represent a static circularly symmetric black hole.
Although the covariant metric components and the electric field do not exhibit
a singular behavior at $r=0$ the curvature invariants are singular at that
point.\\}

{Keywords: 2+1 dimensions, Born-Infeld black hole }\\

PACS numbers: 04.20.Jb

\end{abstract}

\smallskip\

Einstein gravity in (2+1)-dimensions has been intensively studied
in this decade~\cite{Carlip,Mann,Frolov}, largely because of the
existence of black holes solutions in (2+1)-anti de Sitter
spacetimes~\cite{Teitelboim1,Teitelboim2}, which possess certain
features inherent to the (3+1)-black holes. Moreover, it is
believed that (2+1)-gravity will provide new insights towards a
better understanding of the physically relevant (3+1)-gravity.
Nevertheless, to our knowledge, the existing solutions in (2+1)-Einstein
theory, do not consider equations of motion which are 
non-linear in the Maxwell field. Since non-linear electromagnetic
Lagrangians, in particular the Born-Infeld Lagrangian~\cite{Born}, arise in
open string theory (the low-energy effective action for a constant
electromagnetic field is precisely the Born-Infeld
action)~\cite{Fradkin,Tseytlin}, they deserve a special attention. In this 
context, string theory has emerged as the most promising candidate for the
consistent quantization of gravity. In particular, as we mentioned above, the
open string theory has Born-Infeld coupled vector fields, but it is not clear
that this remains the case after compactification to three dimensional space
with negative cosmological constant $\Lambda$. On the other hand, the
Born-Infeld electrodynamics is free from some singularities appearing in the
classical theory of electromagnetic field and one may guess that string
theories with Born-Infeld type effective actions should also be free from
physical singularities .

It should be mentioned that Born-Infeld theory has recently attracted
considerable interest in various contexts~\cite{Larsen,Garcia1}. Among other
notable features, string theory has become a theory that gives interesting
answers towards other fields, such as the physics of black holes, cosmology,
etc.

In (2+1)-dimensions, electromagnetic theories can be constructed
from Lagrangians (without higher derivatives) depending upon a
single (non-vanishing) invariant $F=\frac{1}{4} F_{ab}F^{ab}$,
which can be expressed in terms of the electric (vector ) and
magnetic (scalar) fields: in a lorentzian frame, for an observer moving with
the 3-velocity $v^{a}$, the electric and the magnetic fields are
correspondingly defined as
\begin{eqnarray}
E_{a}=F_{ab}v^{b}, \,\,\, B= \frac{1}{2}
\epsilon_{abc}F_{bc}v^{a},
\end{eqnarray}
where latin indices run the values $0,1,2$ and $\epsilon_{abc}$ is
the totally anti-symmetric Levi-Civita symbol with $\epsilon_{012}=1$, usually
the $v^{a}$ is oriented along the time coordinate, i.e,
$v^{a}=\delta^{a}_{t}$, with such a choice
\begin{eqnarray}
E_{a}=F_{a 0}, \,\,\, B= F_{12}.
\end{eqnarray}
Thus the invariant can be expressed by $F \equiv \frac{1}{4} F^{a b}F_{a b
}=\frac{1}{2}( B^{2}-E^{2})$.

In general one can construct a (2+1)-Einstein theory coupled with
nonlinear electrodynamics starting from the action
\begin{eqnarray}
\label{action} S=\int \sqrt{-g} \left(\frac 1{16\pi} (R-2\Lambda)
+ L(F) \right) \,d^3x,
\end{eqnarray}
with the electromagnetic Lagrangian $L(F)$ unspecified explicitly
at this stage; physically one requires the Lagrangian to coincide
with the Maxwell one at small values of the electromagnetic
fields, $L(F)_{Maxwell}=-F/{4 \pi}$. We are using units in which
$c=G=1$, because of the ambiguity in the definition of the
gravitational constant (there is not newtonian gravitational limit
in (2+1)-dimensions) we prefer to maintain the factor $1/16 \pi$
in the action to keep the parallelism with (3+1)-gravity. Varying
this action with respect to gravitational field gives the Einstein
equations
\begin{eqnarray}
G_{ab} + \Lambda g_{a b}= 8 \pi T_{ab},
\end{eqnarray}
\begin{eqnarray}
T_{ab}=  g_{ab} L(F)- F_{ac} F_{b}^{\,\,c} L_{_{,F}} ,
\end{eqnarray}
while the variation with respect to the electromagnetic potential
$A_{a}$ entering in $F_{ab}= A_{b,a} - A_{a,b}$, yields the
electromagnetic field equations
\begin{eqnarray}
\nabla_{a} \left( F^{ab} L_{_{,F}}  \right)=0.
\end{eqnarray}
We are denoting the derivative of $L(F)$ with respect to $F$ by
$L_{_{,F}}$.
In what follows we shall restrict ourselves to the study of the Born-Infeld
nonlinear electrodynamics with Lagrangian
\begin{eqnarray}
\label{action B-I} L(F)=-\frac{b^{2}}{4 \pi} \left (\sqrt{1+2
\frac{F}{b^{2}}}-1 \right),
\end{eqnarray}
where the constant $b$ is the Born-Infeld parameter. Notice that
this Lagrangian reduce to the Maxwell one in the limit when $b^{2}
\longrightarrow \infty$, $L(F)_{_{Maxwell}}=-F/{4 \pi}$. Therefore
the field equations of the Einstein-Born-Infeld theory amount to
\begin{eqnarray}
G_{ab} + \Lambda g_{ab}= \hspace{3.5cm} \nonumber \\
2 \left(\frac{F_{ac}{F_{b}}^{c}}{\sqrt{1+2F/b^{2}}} - b^{2} g_{ab}
( \sqrt{1+2 \frac{F}{b^{2}}}-1) \right),
\end{eqnarray}
together with the electromagnetic field equations
\begin{eqnarray}
\label{mag}
\nabla_{a} \left ( \frac{ F^{ab}}{\sqrt{1+2 F /b^{2}}}  \right)=0.
\end{eqnarray}

As a concrete solution of Einstein-Born-Infeld dynamical equations
we present a static self-consistent solution. To derive it, we
consider a (2+1)-static circularly metric of the form
\begin{eqnarray}
\label{metrica}
ds^{2}= - f(r) dt^{2} + \frac{dr^{2}}{f(r)} + r^{2} d \Omega^{2},
\end{eqnarray}
where $f(r)$ is an unknown function of the variable r. We restrict the
electric field to be
\begin{eqnarray}
\label{tensorr}
F_{a b} = E(r) \left ( \delta^{t }_{a }  \delta^{r}_{ b} - \delta^{r }_{a}
\delta^{t}_{b} \right ).
\end{eqnarray}
The invariant then is given by
\begin{eqnarray}
\label{invariante}
2 F = -E^{2}(r).
\end{eqnarray}
Substituting~(\ref{tensorr}) and~(\ref{invariante}) into the electromagnetic
field equations~(\ref{mag}) we arrive at
\begin{eqnarray}
\partial_{r} \left (\frac{r E(r)}{\sqrt{1- E^{2}/b^{2}}} \right)=0,
\end{eqnarray}
which integrates as
\begin{eqnarray}
\label{Elec}
E(r)= \frac{q}{\sqrt{r^{2}+ q^{2}/b^{2}}},
\end{eqnarray}
where $q$ is an integration constant having the meaning of the charge, as
one could expect. In the Maxwell limit, we obtain from the last expression the
right $E=q/r$ in (2+1)-dimensions. The Born-Infeld field is characterized by a
charge density distribution $\rho_{e}$, which can be evaluated from the
Maxwell equations $div \vec{E}
= \nabla \vec{E}= 2 \pi \rho_{e}$, which in the considered case amount to
\begin{eqnarray}
div \vec{E} = \frac{1}{r} \frac{d}{dr} \left( r E(r) \right)= 2
\pi \rho_{e},
\end{eqnarray}
substituting here $E(r)$ from~(\ref{Elec}) we obtain
\begin{eqnarray}
\rho_{e} = \frac{q r^{2}_{0}}{2 \pi r (r^{2}+r^{2}_{0})^{3/2}},
\end{eqnarray}
where $r_{0} =q/b $. It is easy to verify that the surface integral
of $\rho_{e}$ is equal to q, in effect
\begin{eqnarray}
\int^{\infty}_{0} \, \rho_{e} dA= q r^{2}_{0} \int^{\infty}_{0} \,
\frac{dr}{(r^{2}+r^{2}_{0})^{3/2}} = q.
\end{eqnarray}
We would like to point out the regular behavior of the vector $\vec{E}$ of the
electric field and the surface charge distribution $\rho_{e}$, the same
regular behavior one encounters for the static spherically symmetric electric
field in (3+1)-Born-Infeld theory.

As far as the Einstein equation are concerned, the $R_{tt}$ and $R_{_{\Omega
\Omega}}$ components yield respectively the equations
\begin{eqnarray}
f_{,rr}+ \frac{f_{,r}}{r}= -4 \Lambda - \frac{4 q^{2}}{r
\sqrt{r^{2} + q^{2}/b^{2}}} \nonumber \\
- 8 b^{2} \left ( \frac{r}{\sqrt{r^{2}+q^{2}/b^{2}}} -1 \right),
\end{eqnarray}
\begin{eqnarray}
f_{,r}= -2 \Lambda r -  \frac{4 q^{2}}{\sqrt{r^{2}+ q^{2}/b^{2}}}
  \nonumber \\
- 4 b^{2}r \left ( \frac{r}{\sqrt{r^{2}+q^{2}/b^{2}}} -1 \right).
\end{eqnarray}
The general integral of this system is given by
\begin{eqnarray}
\label{solucion ultima}
f(r)= - M - (\Lambda - 2  b^{2}) r^{2} -2  b^{2}r
\sqrt{r^{2}+ q^{2}/b^{2}} \nonumber \\ - 2 q^2 ln (r + \sqrt{r^2 + q^2/b^2}).
\end{eqnarray}
From this last expression one sees that there is a contribution of the
Born-Infeld field to the term with the cosmological constant.

Having explicitly the metric one easily calculates the curvature tensor
components:
\begin{eqnarray}
R_{0 1 1 0}=  2 b^{2} - \Lambda - \frac{2  b^{2} r}{\sqrt{r^{2} +
q^{2}/b^{2}}},
\end{eqnarray}
\begin{eqnarray}
R_{0 2 0 2}=  f(r) \left (\Lambda r^{2}- 2 b^{2} r^{2}+
2  b^{2} r \sqrt{r^{2}+ \frac{q^{2}}{b^{2}}} \right)
\end{eqnarray}
and
\begin{eqnarray}
R_{1 2 2 1}=  f(r)^{-1} \left (\Lambda r^{2}- 2 b^{2} r^{2}+
2  b^{2} r \sqrt{r^{2}+ \frac{q^{2}}{b^{2}}} \right).
\end{eqnarray}
It is surprising that the covariant metric and curvature
components do not exhibit a singular behavior in neighborhood of the origin at
$r=0$. (3+1)-static and axially symmetric black holes do not behave in such
a manner. Nevertheless this solution is singular at $r=0$ in the sense that
its invariant characteristics such as the Ricci scalar and the Ricci square
blow up at $r=0$, (the Riemann tensor squared is not necessarily evaluated
since in (2+1)-dimension the Riemann tensor is given in terms of the Ricci
tensor, curvature scalar and the metric tensor). Additionally in
(2+1)-dimension one consider the behavior of the invariant $\det (R_{a b})/\det
(g_{ab})$~\cite{Weinberg}, thus one has to evaluate the invariants
\begin{eqnarray*}
R, \,\,\, R_{ab}R^{ab}, \,\,\, \frac{\det (R_{ab})}{\det (g_{ab})}
\end{eqnarray*}
at critical points. In our case these three invariants are given as
\begin{eqnarray}
R = 6 \Lambda - 12  b^{2} + \frac{4 (2 q^{4} + 5 q^{2} b^{2} r^{2}
+ 3 b^{4}r^{4})}{b^{2} r \left( r^{2}+ q^{2}/ b^{2} \right)^{3/2}},
\end{eqnarray}
\begin{eqnarray}
R_{a b} R^{a b} = \hspace{4cm} \nonumber \\ 3 (4  b^{2}-2 \Lambda)^{2} -
\frac{8  b^{2}(4  b^{2}- 2 \Lambda)}{r}
\left [ \frac{3 r^{2}+ 2
q^{2}/b^{2}}{\sqrt{r^{2}+q^{2}/b^{2}}} \right] \nonumber \\
+ 8 b^{4} \left[2 + \frac{r^{2}}{r^{2}+ q^{2}/b^{2}} + \frac{3
(r^{2}+q^{2}/b^{2})^{2}}{r^{2}}  \right]
\end{eqnarray}
and
%\begin{eqnarray}
%R_{a b c d} R^{a b c d} = \left (4 b^{2} - 2 \Lambda - \frac{4
%b^{2} r}{\sqrt{r^{2}+q^{2}/b^{2}}} \right)^{2}  \nonumber \\ + 4 \left(
% 4 b^{2} - 2 \Lambda - \frac{4  b^{2}}{r} \sqrt{r^{2}+q^{2}/
%b^{2}} \right )^{2}.
%\end{eqnarray}
\begin{eqnarray}
\frac{\det R_{ab}}{\det g_{ab}} = \left(\frac{2 \Lambda - 4  b^{2}}{r^{2}+
\frac{4  b^{2}}{r^{3}} \sqrt{r^{2}+ q^{2}/b^{2}}} \right) \times
\nonumber \\ \left(-2 \Lambda + 4 b^{2} - \frac{2  b^{2} (2 r^{2}+
q^{2}/b^{2})}{r \sqrt{r^{2}+q^{2}/b^{2}}} \right )^{2}.
\end{eqnarray}
Since these scalars go to infinity at $r=0$, we conclude that they are singular
at this point.

\begin{figure}[ht]
\centerline{ \psfig{file=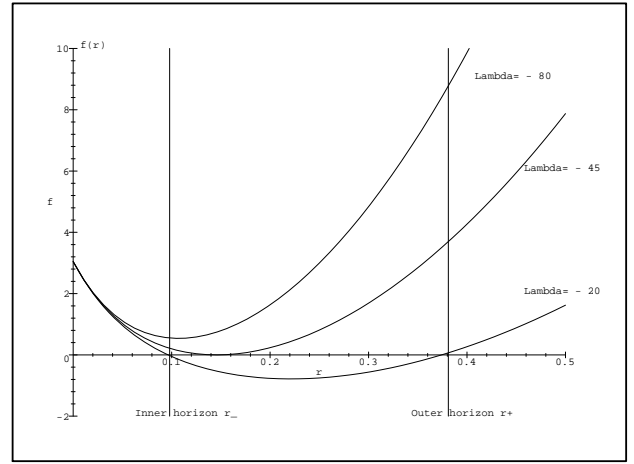,width=9cm,angle=-90} }
\caption{Behavior of $-g_{tt}$ for different values of $\Lambda <
0$.} \label{fig:A}
\end{figure}

This solution is a black hole. To establish this assertion one has
to demonstrate the existence of horizons, which require the
vanishing of the $g_{tt}$ component, i.e., $f(r)=0$. The roots of
this equation give the location of the horizons (inner and outer
in our case). Since this equation is a transcendental one, we are
not able, as it is usual for charged (2+1)-black holes, to express
the roots analytically, even for the charged static BTZ solution,
the roots are expressed in terms of the Lambert $W(x)$ function.
To overcome this difficulty we study the extreme case, in which
the derivative of $\partial_{r} (f(r))=0$ gives
\begin{eqnarray}
\label{rextr} 
r_{extr}= \frac{2 q b}{\sqrt{\Lambda^{2}- 4 b^{2} \Lambda}}  > 0
\end{eqnarray}
for $\Lambda < 0$. Now entering $r_{extr}$ into $f(r)=0$ one obtains a
relation between mass, charge, cosmological constant and the Born-Infeld
parameter, which can be solved explicitly for the mass--the extreme one--
\begin{eqnarray}
M_{extr}=- 2 q^{2} \, ln \, \left( \frac{q}{b} \sqrt{\frac{\Lambda- 4
b^{2}} {\Lambda}}  \right).
\end{eqnarray}
We have an extreme black hole if $\Lambda < 0$, $M_{extr} > 0$ and
$q^{2} < b^{2}$, this last constraint arises from $\Lambda < 4
b^{2} q^{2}/(q^{2}- b^{2})$ when one demands $M_{extr}> 0$. Fixing
the values of the $M_{extr}$ for given values of $q$, $b$ and
$\Lambda$ one has a black hole solution with inner and outer
horizons when $M > M_{extr}$. For $M < M_{extr}$ one has a soliton
solution, i.e., there are no horizons at all and we have a naked
singularity (see FIG.~(\ref{fig:A})). It may occur that for certain values of
the parameters there is only one positive root of the equation
$f(r)=0$ -- the horizon $r_{h} > 0$ -- in such case one has also a
black hole solution.  A similar analysis can be
carried out for other $r_{extr}$ which arises for $\Lambda > 0$, this
important case will be treated elsewhere.

At infinity, for weak electromagnetic field our solution
asymptotically behaves as the charged
BTZ~\cite{Teitelboim1,Teitelboim2}:
\begin{eqnarray}
ds^2= - \left( - M + \frac{r^{2}}{l^{2}} - q^{2}_{e} ln \, r \right)
dt^2 \nonumber \\
+ \frac{dr^{2}}{ \left(- M + \frac{r^{2}}{l^{2}} - q^{2}_{e}
ln \, r \right)} + r^2 d\phi^2,  \label{BTZ}
\end{eqnarray}
and
\begin{eqnarray}
F_{a b} = \frac{q}{r} \left ( \delta^{t }_{a }  \delta^{r}_{ b} - \delta^{r }_{a}
\delta^{t}_{b} \right ).
\end{eqnarray}
with mass $M$, cosmological constant $\Lambda= -1/l^{2}$, and
charge $q$. Hence for $\Lambda=- 1/l^{2}$ our solution at infinity
behaves as anti-de Sitter spacetime. As can be seen directly from
this  BTZ metric, it is singular at $r=0$. If one require
additionally the vanishing of the cosmological constant one
arrives at a solution reported in~\cite{Gott}.

As regards the analytical extension of our solution, one has to
follow step by step the procedure presented in the standard
textbooks (for instance~\cite{Wald}) to determine the
Kruskal-Szekeres coordinates. First one has to integrate for the
tortoise $r_{*}$ coordinate: $r_{*}= \int{1/f(r)}dr$, which in our
case has no expression in terms of elementary functions, next one
defines the null coordinate $u$ and $v$ by $u=t-r_{*}$,
$v=t+r_{*}$; in these coordinates the studied metric acquires the
form
\begin{eqnarray}
ds^2= -f(r) du dv + r^2 d \Omega^2,
\end{eqnarray}
where $r$ has to be interpreted as function of $u$ and $v$,
$r_{*}=r_{*}(u,v)$. Further, one introduces null Kruskal-Szekeres
coordinates $U= - e^{- \alpha u}$ and $V= e^{\beta V}$ where
$\alpha$ and $\beta$ to be chosen appropriately, finally one
introduces the Kruskal-Szekeres coordinates $T=(U+V)/2$, X=
(V-U)/2, arriving at the Kruskal extension.

If one were interested in the thermodynamics of the obtained
solution one would to evaluate the temperature of the black hole, which is
given in terms of its surface gravity by~\cite{Visser,Brown}
\begin{eqnarray}
\label{T}
k_{_{B}} T_{_{H}} = \frac{\hbar}{2 \pi} \, k.
\end{eqnarray}
In general, for a spherically symmetric (and for circularly symmetric in
(2+1)-dimensions) system the surface gravity can be computed via (for our
signature)
\begin{eqnarray}
\label{k}
k=- \lim_{r \rightarrow r_{_{+}}} \left [\frac{1}{2} \frac{\partial_{r}
g_{tt}}{\sqrt{- g_{tt} g_{rr}}}  \right ],
\end{eqnarray}
where $r_{_{+}}$ is the outermost horizon. For our solution we have
from~(\ref{solucion ultima}),~(\ref{T}) and~(\ref{k}) that
\begin{eqnarray}
\label{temp}
k_{_{B}} T  = \frac{\hbar}{4 \pi} \left(-2 (\Lambda - 2 b^2) r_{_{+}} - 4 b^2
\sqrt{r_{_{+}}^2 + \frac{q^2}{b^2}} \right ).
\end{eqnarray}
Since in our case there is no an analytical expression of $r_{_{+}}$ in terms
of elementary functions, one can not give a parameter dependent expression 
of~(\ref{temp}). It is easy to check that when $q=0$, $T$ in~(\ref{temp})
reduces to the BTZ temperature. In the extremal case~(\ref{rextr}), the
temperature vanishes in~(\ref{temp}). The entropy can be trivially obtained
using the entropy formula $S= 4 \pi r_{_{+}}$. Other thermodynamic quantities
such as heat capacity and chemical potential can be computed as
in~\cite{Brown}. We recall that most of these quantities in the literature are
evaluated for metrics given in terms of polynomial functions.

Notice that the four dimensional Einstein-Born-Infeld counterpart-- the
Kottler-Born-Infeld black hole~\cite{Garcia2,Gibbons}-- can be given by the
metric~(\ref{metrica}) with
\begin{eqnarray}
f(r) = 1-2M/r- (\Lambda/3- 2  b^{2}/3)r^{2} \nonumber \\ -
\frac{2}{3} b^{2} \sqrt{r^{4}+q^{2}/b^{2}} - 2  q^{2}/3r
\int\frac{dr}{\sqrt{r^{4}+q^{2}/b^{2}}}
\end{eqnarray}
where now $d \Omega^{2}= d\theta^{2}+sin^{2}\theta d \phi^{2}$. As in the
(2+1)-case, here there is also a contribution to the cosmological constant
term of the nonlinear field. The corresponding electric field is given by
\begin{eqnarray}
E(r)=\frac{q}{\sqrt{r^{4}+ q^{2}/b^{2}}}.
\end{eqnarray}
Notice that the electric field in this case is regular everywhere. This
gravitational field asymptotically behaves as the Kottler charged solution,
with the structural function and electromagnetic field of the form
\begin{eqnarray*}
f(r) = 1- \frac{2 M}{r} - \frac{\Lambda}{3} r^{2} +
\frac{q^{2}}{r^{2}}+ O \left (\frac{1}{r^{6}} \right),
\\ E (r)=  \frac{q}{r^{2}}+ O \left( \frac{1}{r^{3}} \right)
\hspace{3cm}.
\end{eqnarray*}
By cancelling $\Lambda$ one obtains an asymptotically flat solution.

This work was supported in part by FONDECYT-Chile 1990601, Direcci\'{o}n de
Promoci\'{o}n y Desarrollo de la Universidad del B\'{\i}o-B\'{\i}o through
Grant No 983105-1 (M.C.); FONDECYT-Chile 1980891, CONACYT-M\'exico
3692P-ER9607 (A.G.) and in part by Dicyt (Universidad de Santiago de
Chile) (M.C., A.G.).


\begin{references}
\bibitem{Carlip} S. Carlip, ``Lectures on (2+1)-dimensional Gravity", Davis
preprint UCD-95-6, gr-qc/9503024.

\bibitem{Mann} R. Mann, ``Lower Dimensional Black Holes: Inside and out",
qr-qc/9501038

\bibitem{Frolov}  V. Frolov, S. Hendy and A.L. Larsen, Nucl. Phys. B {\bf 468}
 (1996) 336.

\bibitem{Teitelboim1}  M. Ba\~nados, C. Teitelboim and J. Zanelli, Phys.
Rev. Lett. {\bf 69} (1992) 1849.

\bibitem{Teitelboim2}  M. Ba\~nados, M. Henneaux, C. Teitelboim and J.
Zanelli, Phys. Rev. D {\bf 48} (1993) 1506.

\bibitem{Born} M. Born and L. Infeld, Proc. Roy. Soc. (London)
{\bf A 144} (1934) 425.

\bibitem{Fradkin} E. Fradkin and A. Tseytlin, Phys. Lett B {\bf
163} (1985) 123.

\bibitem{Tseytlin} A. Tseytlin, Nucl. Phys. {\bf 276} (1986) 391.

\bibitem{Larsen} A.L. Larsen, N. S\'anchez, Strings and Multi-Strings in Black
Holes and Cosmological Spacetimes, in: New Developments in String Gravity and
Physics at the Planck Energy Scale, World Scientific, 1995.

\bibitem{Loll} R. Loll, J. Math. Phys. {\bf 36} (1995) 6494.

\bibitem{Deser}  S. Deser and G.W. Gibbons, Class. Quant. Grav. {\bf 15}
 (1998) L35.

\bibitem{Dariusz}  D. Chruscinski, Phys. Lett. A {\bf 240} (1998) 8.

\bibitem{Callan}  C.G. Callan and J.M. Maldacena, Nucl. Phys. B {\bf 513}
 (1998) 198.

 \bibitem{Garcia1} H. Salazar, A. Garc\'\i a and J. Plebanski, J. Math. Phys.
{\bf 28} (1987) 2171.

\bibitem{Weinberg} S. Weinberg, Gravitation and Cosmology (John Wiley $\&$
Sons, New York, 1972)

\bibitem{Gott} J.R. Gott III, J.Z. Simon and M. Alpert, Gen. Rel. Grav. {\bf
18} (1986) 1019.

\bibitem{Wald} R. Wald, General Relativity (The University Chicago Press,
1984).

\bibitem{Garcia2} H. Salazar, A. Garc\'\i a and J. Plebanski, Nuovo Cimento
{\bf B 84} (1984) 65.

\bibitem{Visser} M. Visser, Phys. Rev. D {\bf 46} (1992) 2445.

\bibitem{Brown} J.D. Brown, J. Creighton and R. Mann, Phys. Rev. D {\bf 50} 
(1994) 6394.

\bibitem{Gibbons}  G.W. Gibbons and D.A. Rasheed, Nucl. Phys. B {\bf 454}
 (1995) 185.
\end{references}
\end{document}